\documentclass[12pt,preprint]{aastex}

\newcommand{\OI}{O\,{\sc i}}
\newcommand{\CII}{C\,{\sc ii}}
\newcommand{\emm}[1]{\ensuremath{#1}}   
\newcommand{\emr}[1]{\emm{\mathrm{#1}}} 

\slugcomment{ApJ Letters;
\textit{Received 2009 May 14; accepted 2009 June 2}}

\shorttitle{Detection of neutral atomic oxygen
toward the  Horsehead}
\shortauthors{Goicoechea et al.}

\begin{document}


\title{Far-Infrared detection of neutral atomic oxygen\\
toward the  Horsehead Nebula$^{1}$}


\author{Javier R. Goicoechea}
\affil{Centro de Astrobiolog\'{\i}a (CSIC-INTA), Laboratorio de Astrof\'{\i}sica Molecular, \\
Carretera de Ajalvir, Km 4. Torrej\'on de Ardoz, 28850 Madrid, Spain.}
\email{goicoechea@damir.iem.csic.es}

\author{Mathieu Compi\`egne}
\affil{Canadian Institute for Theoretical Astrophysics, 
University of Toronto, \\ 60 St. George Street, Toronto, ON  M5S 3H8, Canada}
\email{compiegne@cita.utoronto.ca}

\and

\author{Emilie Habart}
\affil{Institut d'Astrophysique Spatiale, 
Universit\'e Paris-Sud, 91405 Orsay Cedex, France.}
\email{emilie.habart@ias.u-psud.fr}


\altaffiltext{1}{This work is based on observations made with the 
\textit{Spitzer Space Telescope}, 
which is operated by the Jet Propulsion Laboratory, 
California Institute of Technology under a contract with NASA.}


\begin{abstract}
We present the first detection of neutral atomic oxygen ($^3P_1$--$^3P_2$  
fine structure line at $\sim$63\,$\mu$m)
toward the Horsehead photodissociation region (PDR). The cloud has
been mapped  with the  \textit{Spitzer Space Telescope} at far-IR (FIR) wavelengths 
using the Multiband Imaging Photometer for \textit{Spitzer} (MIPS)
in the spectral energy distribution (SED) mode.
The [\OI]63\,$\mu$m line peaks at the illuminated edge of the cloud at $A_V$$\simeq$0.1-0.5
(inward the gas becomes too cold and outward the gas density drops).
The luminosity carried by the  [\OI]63\,$\mu$m line represents a significant fraction of 
the total FIR dust luminosity ($I_{63}$/$I_\emr{FIR}$$\simeq$4$\times$10$^{-3}$).
We analyze the dust continuum emission and the \textit{nonlocal} \OI\, excitation and
radiative transfer in detail. The observations are reproduced with
a gas density of $n_\emr{H}$$\simeq$10$^4$\,cm$^{-3}$ and 
gas and dust temperatures of $T_k$$\simeq$100\,K and $T_d$$\simeq$30\,K.
We conclude that the determination of the \OI~$^3P_J$ level populations and emergent line intensities
at such ``low'' densities is a complex non-LTE problem. FIR radiative pumping, 
[\OI]63\,$\mu$m~subthermal emission, 
[\OI]145\,$\mu$m suprathermal and even  maser emission can occur and decrease  the
resulting  [\OI]63/145 intensity ratio. 
The \textit{Herschel Space Observatory}, observing from $\sim$55 to 672\,$\mu$m,  
will allow us to  exploit the diagnostic power
of FIR fine structure lines with unprecedented resolution and sensitivity.

\end{abstract}



\section{Introduction}

Oxygen is the most abundant heavy element in the 
Universe. Due to its relatively high ionization potential, far-ultraviolet (FUV)
photons (6\,eV$<$$h\nu$$<$13.6\,eV) do not ionize it and thus a large fraction
of atomic oxygen can be expected in the photodissociated gas interface
between the ionized and the molecular gas before chemical  species such
as CO, H$_2$O or OH reach high abundances.
The absorption of FUV photons by dust grains 
leads to the ejection of photoelectrons that heat the gas (\textit{e.g.,} Draine 1978)
as well as to  intense FIR dust continuum emission. The efficiency of this essential process
can be estimated by comparing the luminosity of the main gas cooling lines (\textit{e.g.,}  
the FIR fine structure lines of \CII\, and \OI) with the luminosity of the  dust emission.
Indeed,  FIR  lines from abundant gas reservoirs  
are key players in the energy balance of the regions they trace (from FUV illuminated
interstellar clouds and protoplanetary disk surfaces  to extragalactic starbursts).

The \OI($^3P_J$) fine structure triplet system is formed by the $^3P_2$ (ground--state),
the $^3P_1$~($\Delta$$E$$_{12}$/$k$=228\,K) and the  $^3P_0$   ($\Delta$$E$$_{01}$/$k$=99\,K) levels.
Radiative transitions between the $^3P_1$--$^3P_2$ and $^3P_0$--$^3P_1$ states lead
to the widespread [\OI]63 
and 145\,$\mu$m line emission respectively. Observations and models
of such lines are crucial diagnostics of  a broad
range of galactic and extragalactic environments 
(see  Hollenbach \& Tielens 1999 for a review).

In this Letter, we present the first detection of the [\OI]63\,$\mu$m line toward the
Horsehead PDR ($d$$\sim$400\,pc, edge-on), one of the most famous FUV-induced chemical laboratories in the sky.
We analyze the FIR line and continuum emission without strong simplifying assumptions
in the excitation and radiative transfer treatment.

\section{Observations and Data reduction}

The Horsehead nebula was observed with the Multiband Imaging Photometer for \textit{Spitzer} 
(MIPS) as part of our ``SPECPDR'' program (ID:20281; Joblin et al.~2005) 
on 2006 April~3 using the spectral energy distribution 
mode (hereafter MIPS/SED).
MIPS/SED provided the capability of obtaining low--resolution spectra ($R\,=\,15-25$) in the FIR
range ($\lambda=55-97\,\mu$m)
using a slit  spectrometer and the MIPS\,70 Ge:Ga detector array. 
The slit size has two detector pixels 
in width ($\sim$20$''$) and 12 pixels  in length ($\sim$2$'$).
A reflective grating provides a spectral dispersion of 
1.7\,$\mu$m pixel$^{-1}$.
We used the ``spectral mapping'' mode and covered the Horsehead
with 14 pointings shifted by half the slit width
in the direction perpendicular to the slit length. 
The exposure time was 10\,s per pointing.
The ``chop distance'' for the OFF positions was set to +3$'$ for each pointing.
Both ON-- and OFF--source areas are shown in Fig.\,\ref{fig:show_area}$a$.

We subtract the average OFF-source spectrum to every ON-source spectra.
We first retrieved the BCD data (ver. s16.1.0) that we proceed with 
the {\it mosaic\_sed.pl} script of 
MOPEX\footnote{http://ssc.spitzer.caltech.edu/postbcd/mopex.html} to obtain the 
post-BCD level data (in intensity units of MJy\,sr$^{-1}$).
We apply a flat correction to the spatial dimension on the 
detector (\textit{i.e.} detector columns relative gain correction) at the BCD level.
We compute the flat response of the detector columns using the OFF positions. 
We have developed a pipeline that builds spectral cubes (two spatial dimensions
and one spectral dimension) from the Post-BCD data.
We work in the assumption that no spatial information is available in the slit width.
Thus, the $N_\lambda$ rows of the detector can be seen as images of the sky through the slit 
for $N_\lambda$ sampled wavelengths  between 55 and 97\,$\mu$m.
For each wavelength, this image of the slit is reprojected into a 
$N_\alpha\times N_\delta$ spatial grid of the sky.
The final cube then has $N_\alpha \times N_\delta\times N_\lambda$ pixels.
The pixel size of the spatial grid is half of each detector pixel field--of--view, 
PFOV (9.8''/2 = 4.9'').

Nevertheless,  due to  dispersion,
the light that falls on a pixel comes from the entire slit
width, that is two times the pixel size, so that
the PFOV is rather $9.8"\times 19.6"$. This leads to an overestimation
of the intensity of a factor 2 that we take into account here. 
For extended sources, one also has to correct for the effect of the aperture
used for the photometric calibration with point sources. We then divide 
the intensity by the 5 pixels aperture correction function (see MIPS Data Handbook).
The photometry of the resulting  spectral cubes (for all sources in our program)
were checked by cross-calibrating them with MIPS\,70 data when available.
It appears that the MIPS/SED gain is 30\% higher than the MIPS\,70 one.
On the basis that the photometer calibration is more reliable, we apply this 30\% correction 
to our MIPS/SED data and use the MIPS\,70 absolute
calibration uncertainty of 7\%.\\

\section{Observational Results}

\begin{figure*}[ht]
\centering %
\includegraphics[width=1.1\textwidth,angle=0]{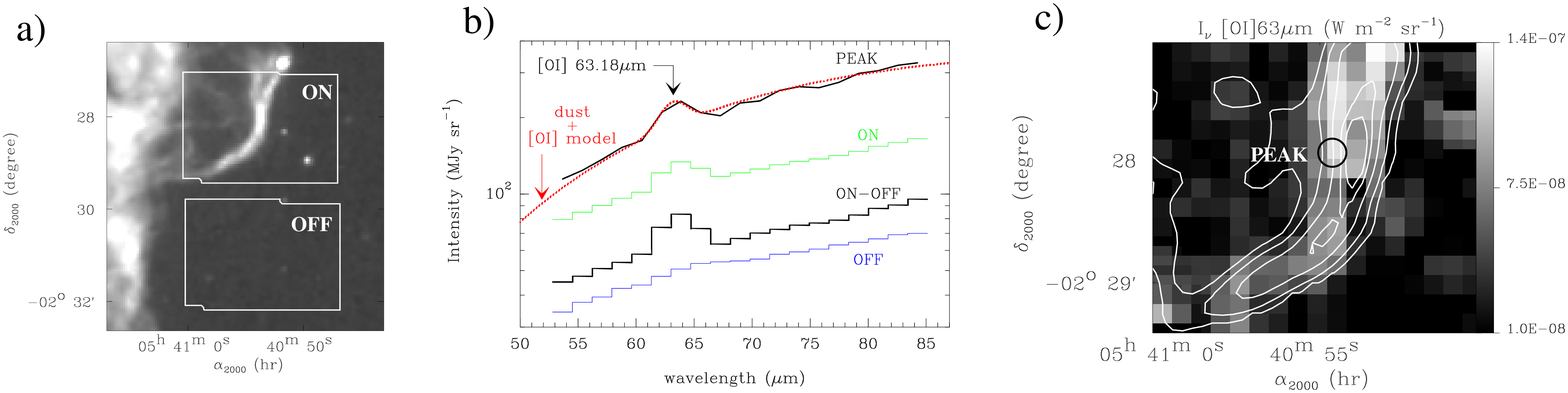} 
\caption{\textit{(a)} ISOCAM (5-8.5\,$\mu$m) image of the Horsehead
 (dominated by PAH emission; Abergel et al.~2003). 
The ON-- and OFF--source areas of our \textit{Spitzer}/MIPS observations are shown. 
\textit{(b)} Different spectra:
mean ON--source (green histogram), OFF--source (blue histogram),
and ON--OFF (black histogram). The MIPS spectrum toward the PDR
(PEAK position) is shown on top (black continuous curve) together with our dust and [\OI]
radiative transfer model convolved with MIPS/SED spectral resolution. 
\textit{(c)} [\OI]63\,$\mu$m  intensity map. The scale
 is shown in the look-up table (in W\,m$^{-2}$\,sr$^{-1}$). The white contours
represent the  \textit{ortho}-H$_2$~$v$=0--0 $S$(1) line intensity at $\sim$17\,$\mu$m 
(0.9$\times$, 1.5$\times$, 2.5$\times$, 3.5$\times$, and 4.0$\times$10$^{-8}$\,W\,m$^{-2}$\,sr$^{-1}$;
Habart et al. 2009). 
The circle represents the $\sim$15$''$-FWHM resolution toward the PEAK.}
\label{fig:show_area}
\end{figure*}

Figure\,\ref{fig:show_area}$a$  shows the ON-- and OFF--source areas of our \textit{Spitzer}
observations plotted over the polycyclic aromatic hydrocarbon (PAH) mid--IR  
emission toward the Horsehead  (Abergel et al. 2003). 
The PAH band emission traces the FUV illuminated edge of the cloud
(Compi\`egne et al. 2007, 2008; Goicoechea et al. 2009).
MIPS spectra averaged over the ON and OFF  areas are shown
as green and blue histograms (Fig.\,\ref{fig:show_area}$b$). 
The ON--OFF subtracted spectrum is shown as a black histogram.
The [\OI]63\,$\mu$m line is not present in the OFF spectrum, and therefore,
the extended line emission  can fully be attributed to the cloud. 

In order to have an approximate view of the spatial location of the oxygen emission, 
Fig.\,\ref{fig:show_area}$c$ displays the [\OI]63\,$\mu$m line intensity
measured in each pixel of the data cube 
(gray scale). The \textit{ortho}-H$_2$ $v$=0--0 $S$(1) pure rotational line emission at 
$\sim$17\,$\mu$m, delineating the edge of the PDR, 
is  shown as white contours. Overall, the [\OI] emission
peaks in the warm cloud layers (where H$_2$ and PAH emission are also bright) 
directly exposed to the FUV radiation from $\sigma$Ori O9.5V star.
Although the excitation  requirements of the \textit{ortho}-H$_2$ $v$=0--0 $S$(1)  
($E$$_u$/$k \simeq$1015\,K) and 
[\OI]63\,$\mu$m ($E$$_u$/$k \simeq$228\,K) lines are different,
the apparent shift between both line  emission peaks 
in Fig.\,1$c$ may be marginal since
MIPS/SED astrometry is not better than $\sim$5$''$. Besides, 
cloud inclination variations (relative to the line of sight) from north to south,
and thus differential line opacity effects, likely explain 
the brighter  [\OI]63\,$\mu$m emission observed toward the north of the filament. 

The [\OI]63\,$\mu$m line intensity toward the emission peak, as
measured over the $\sim$15'' full width at half maximum (FWHM)
 angular resolution of MIPS at 63\,$\mu$m 
 (circle on Fig.\,\ref{fig:show_area}$c$), is 
$I_{63}$=(1.04$\pm$0.14)$\times$10$^{-7}$~W\,m$^{-2}$\,sr$^{-1}$
($\sim$4 times brighter than the H$_2$ $v$=1--0 $S$(1) fluorescent emission
at 2.12\,$\mu$m; 
see Habart et al. 2005). 
The resulting MIPS spectrum
is shown in Fig.\,\ref{fig:show_area}$b$ as ``PEAK''. 
The [\OI]63\,$\mu$m line is detected at a 10$\sigma$ level.
The FIR thermal emission
of dust raises from $\sim$155\,MJy~sr$^{-1}$ (at $\sim$55\,$\mu$m) to 
$\sim$325\,MJy~sr$^{-1}$ (at $\sim$85\,$\mu$m). The continuum peaks
at longer wavelengths and thus the temperature of the grains emitting
in the FIR  ($T_d$) cannot be high ($T_d$$\lesssim$35\,K), lower than
the expected gas temperature ($T_k$).
Apart from this [\OI]63\,$\mu$m bright emission toward the cloud edge,
the MIPS data show a fainter level of widespread   emission
($\sim$4 times weaker) likely arising from the more diffuse cloud envelope,
or halo, illuminated by the ambient FUV radiation field.

\section{Discussion}

\subsection{Physical conditions in the Horsehead PDR}

In order to guide our interpretation of the [\OI]63$\mu$m line  
toward the PEAK, 
Fig.\,\ref{fig:mtc_grid} (\textit{left}) shows the output of a photochemical model
(Le Petit et al. 2006; Goicoechea \& Le Bourlot~2007)
adapted to the steep density gradient,  $n_\emr{H}$$\sim$$r^{-3}$ (Habart et al. 2005, Goicoechea et al. 2009) 
and FUV illumination conditions in the Horsehead edge  ($G_0$$\simeq$100, where
$G_0$ is the FUV flux in Habing units of the local interstellar value).
We adopt an elemental oxygen   abundance of 3$\times$10$^{-4}$ relative
to H nuclei (Savage \& Sembach 1996).
Taking into account that the gas temperature rapidly decreases from the low density
PDR ($T_k$$\simeq$100\,K; see also Pety et al. 2005; Habart et al. 2009) 
to the denser and shielded inner layers 
($T_k$$\lesssim$20\,K; Pety et al. 2007), we predict that the  
[\OI]63$\mu$m emission peak should occur at the surface of the molecular cloud  between
$A_V$$\simeq$0.1 and~0.5. Due to the relatively low FUV radiation field impinging
the cloud, dust grains are not expected to  attain large temperatures
($T_d$$\lesssim$30\,K). However, H$_2$ photodissociation  is still 
important in the cloud edge and a significant fraction of hydrogen is predicted to 
reside in atomic form in the same layers where [\OI]63\,$\mu$m is bright.
At extinction depths greater than $A_V$$\sim$1.5, carbon monoxide starts to
be protected against 
photodissociation and the gas temperature falls sharply.
The formation of CO (and other O-rich
gas and ice species) decreases the abundance of atomic oxygen inward.

\subsection{Nonlocal and Non-LTE Modeling}

Observations of the [\OI]63\,$\mu$m line and FIR continuum emission
constrain the mean gas density and both the 
dust and gas temperatures toward the PDR.
In this work, we use a \textit{nonlocal} and non-LTE code that treats both the 
line and  continuum radiative transfer  and includes the radiative pumping
of \OI\,~levels by absorption of FIR dust photons (see the Appendix in Goicoechea et al. 2006).
Almost all studies solving the excitation of atomic  fine structure levels simplify the
problem by assuming a \textit{local} treatment of the excitation 
calculation:  \textit{LVG} or single-zone escape probability
methods (\textit{e.g.,} R\"ollig et al. 2007), where the source function 
(the line emissivity/absorption ratio) is assumed to be constant,
and the radiative coupling between different cloud positions is neglected.
The effects of dust opacity are also generally omitted (Gonz\'alez Garc\'{\i}a et al. 2008).
Their applications to the interpretation of [\OI]63,145\,$\mu$m lines
have been widely used by a number of authors in the past (\textit{e.g.,}~Tielens \& Hollenbach 1985; 
van Dishoeck \& Black 1986; Spaans et al. 1994; Poglitsch et al. 1995; 
Kaufman et al. 1999; Vastel et al. 2001; Liseau et al. 1999, 2006). 
Nevertheless, Elitzur \& Asensio--Ramos (2006) recently compared the exact solutions provided
by their \textit{nonlocal} method with those given by the \textit{local} approximation and 
concluded that, for gas densities around 10$^3$--10$^4$\,cm$^{-3}$, 
the error in the predicted [\OI]63/145 line intensity ratio can be
significant (\textit{i.e.,}~comparable with the variation range of the
ratio for different temperatures and densities).
Indeed, these are the densities
expected in the outer layers of the Horsehead nebula, \textit{i.e.,} 
 below the critical densities of \OI\, fine structure  transitions
 ($n_\emr{cr}$$\simeq$5$\times$10$^5$\,cm$^{-3}$ for 
\OI\,($^3P_1$--$^3P_2$)--H$_2$ collisional\footnote{The collisional rate
coefficients ($\gamma_{ij}$) used in this work are:
\OI\,~collisions with \textit{ortho/para}--H$_2$ from Jaquet et al. (1992):
note that for low energies ($T_k$$\lesssim$300-400\,K),  the
$\gamma_{02}$ and $\gamma_{12}$ rates are 
slightly higher (\textit{e.g.,} by $\lesssim$10$\%$ at 100\,K) when \textit{para}-H$_2$ is the perturber. 
Otherwise, rate coefficients are slightly higher for \textit{ortho}-H$_2$ collisions;
\OI--H collisions from Abrahamsson et al. (2007): at 100~K these excitation 
rate coefficients are a factor of $\sim$4 larger than those by  Launay \& Roueff (1977); 
\OI--He collisions from Monteiro \& Flower (1987);
\OI--$e^-$ collisions from Bell et al. (1998) and
\OI--$p^+$ collisions from Chambaud et al. (1980).} de-excitation).
Under these conditions, the determination of the  $^3P_J$ level populations 
is a non-trivial problem and radiative and opacity effects 
(FIR pumping and line-trapping) can play a  role.
Besides, the [\OI]63\,$\mu$m line becomes optically thick ($\tau_{63}>1$) for 
low  column densities, $N$(\OI)$\gtrsim$10$^{17}$\,cm$^{-2}$,
and the  [\OI]145\,$\mu$m line can be suprathermal ($T_{ex}$$>$T$_k$) or
even inverted ($T_{ex}$$<$0) and the emission amplified ($\tau_{145}$$<$0) 
by the maser effect (see also Liseau et al. 2006; Elitzur \& Asensio--Ramos 2006).

\subsection{FIR dust continuum emission}

In order to constrain the dust temperature, the FIR dust luminosity
and obtain  an approximated value for the gas density in the [\OI]63\,$\mu$m
emitting region, we have first modeled the observed SED  toward the PEAK position. 
We explicitly solve the continuum radiative transfer problem  by taking
a dust opacity $\tau_d(\lambda)$  proportional to $\kappa_{100}$(100/$\lambda$)$^{\beta}$,
where $\kappa_{100}$ is the dust opacity at 100\,$\mu$m (per gas+dust mass column density)
and $\beta$ is the dust spectral index.
Assuming a line-of-sight depth of 0.1\,pc (uncertain within a factor of $\sim$2,
see constraints by Habart et al. 2005), 
the best models of the continuum shape and absolute intensity level are 
obtained for $T_d$$\simeq$30\,K (in agreement with the PDR
model prediction) and $n_\emr{H}$=$n$(H)+2$n$(H$_2$)$\simeq$10$^{4}$\,cm$^{-3}$  (with
$\beta$$\simeq$1.4 and $\kappa_{100}$$\simeq$0.25\,cm$^2$\,g$^{-1}$; 
\textit{e.g.,}~Li \& Draine 2001). Integrating the resulting 
continuum intensity distribution from 40 to 500\,$\mu$m we 
obtain $I_\emr{{FIR}}$=2.5$\times$10$^{-5}$
W\,m$^{-2}$\,sr$^{-1}$. 
The dust  model is shown in 
Fig.\,\ref{fig:show_area}$b$ (red dashed curve). 
We determine that the line-of-sight extinction
depth toward the PEAK is $A_V$$\simeq$1.5, consistent with the expected
values of a cloud edge. The FIR dust continuum toward the region
of bright [\OI], H$_2$, and PAH emission is optically thin, with 
$\tau_d(63)$$\simeq$2$\times$10$^{-3}$. 

\subsection{\OI\, excitation and radiative transfer}

After modeling the FIR radiation field  in the PDR we now solve the
\OI\,  $^3P_J$  populations adding radiative pumping
by FIR dust photons. We include 
collisional rate coefficients$^2$ (cm$^3$\,s$^{-1}$) of atomic oxygen with
\textit{ortho}-- and  \textit{para}--H$_2$, H, He, 
electrons, and protons.
H$_2$ and H collisions dominate, but it is important to include their rates (s$^{-1}$)
with the most realistic H and H$_2$ density contributions. 
Note that, for example, $^3P_1-^3P_0$
collisional excitation can be neglected (to first order) 
if only \OI--H$_2$ collisions apply (Monteiro \& Flower 1987), 
but it does contribute if collisions with H are included (Abrahamsson et al. 2007). 
We adopt an \textit{ortho}-to-\textit{para}--H$_2$  ratio$^2$ (OTP) of 3, although, 
at the $T_k$ and $n_\emr{H}$ considered in this work, our results do not change 
by taking an OTP of 1 ([\OI] lines become $\sim$2$\%$ brighter).

\begin{figure*}[ht]
\centering %
\includegraphics[width=1\textwidth,angle=0]{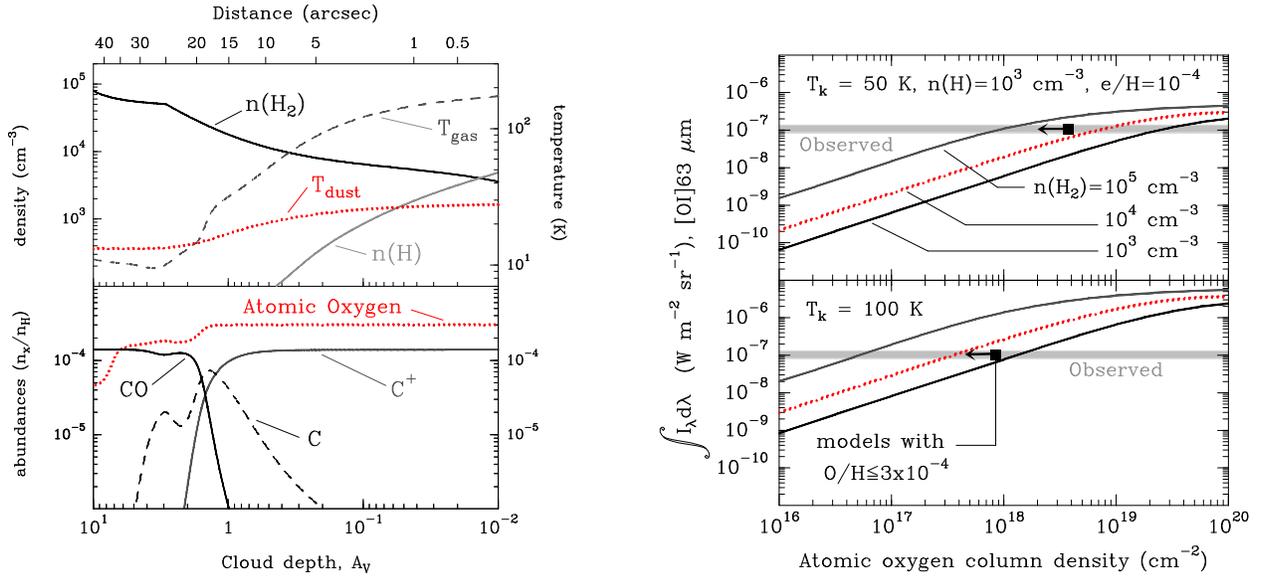} %
\caption{\textit{Left}: PDR model adapted to the Horsehead ($G_0$=100 and density gradient 
from Goicoechea et al. 2009). The upper panel shows the
resulting H$_2$ and H density profiles, together with the $T_k$ and $T_d$
 temperature gradients. The lower panel shows the CO/C/C$^+$ transition and the predicted
abundance profile of atomic oxygen. 
\textit{Right}: output of a grid of \textit{nonlocal}, non-LTE radiative transfer
models. Panels show the expected [\OI]63\,$\mu$m line
intensity (in W\,m$^{-2}$\,sr$^{-1}$) as a function of $N$(\OI) for two gas temperatures 
($T_k$=50 and 100\,K) and different gas densities: $n$(H)=10$^3$\,cm$^{-3}$ fixed and 
varying $n$(H$_2$). The shaded regions show the  [\OI]63\,$\mu$m line intensity
observed by \textit{Spitzer}/MIPS toward the PEAK.} 
\label{fig:mtc_grid}
\end{figure*}

Figure~\ref{fig:mtc_grid} (right) displays a grid of \textit{nonlocal} 
models showing the emergent $I_{63}$ intensity as a function of $N$(\OI)
column density. The lower panel shows the predicted $I_{63}$ for
$T_k$=100\,K (\textit{i.e.,}~roughly the gas temperature in the outermost 
cloud layers) and varying
H$_2$ densities. 
Guided by our PDR model (Figure~\ref{fig:mtc_grid} (left)), and
in order to modify only one physical parameter in the grid,
we take a fixed $n$(H) density of 10$^{3}$\,cm$^{-3}$ and a high ionization fraction:
$n(e^-)$/$n_\emr{H}$=10$^{-4}$ (see  Goicoechea et al. 2009).
Our calculations include thermal, turbulent, and opacity line broadening
with a turbulent velocity dispersion of $\sigma$=0.225\,km\,s$^{-1}$\,
(FWHM$=2.355\times\sigma$). This value is
inferred from heterodyne interferometric observations of the HCO radical toward 
the PDR (Gerin et al. 2009).
The shaded regions show the observed  [\OI]63\,$\mu$m line intensity 
toward the PEAK position. For a line-of-sight depth of 0.1\,pc 
and taking \OI/H$\simeq$3$\times$10$^{-4}$ in the PDR (most oxygen in atomic form), 
our observations are reproduced with $n$(H$_2$)$\simeq$5$\times$10$^3$\,cm$^{-3}$
(consistent with the dust model)
and $N$(\OI)$\simeq$8.5$\times$10$^{17}$\,cm$^{-2}$
(best model  shown in Figure~1 (b)). The [\OI]63\,$\mu$m 
opacity at line center is $\tau_{63}$$\simeq$6 
(due to saturation the width of the line profile is opacity broadened, 
FWHM$\sim$1.3\,km\,s$^{-1}$, 
\textit{i.e., } broader than the linewidth of an optically thin line affected only
by  turbulent and thermal dispersion). 
The predicted  
[\OI]63/145 line intensity ratio is $\simeq$15.
The required H$_2$ density has to be multiplied~(divided) by $\sim$4 if the gas
temperature is a factor of 2 lower~(higher); line opacities and [\OI]63/145 intensity ratios change
accordingly. Future observations of both [\OI]63,145\,$\mu$m lines will help to 
better constrain the density and the atomic oxygen abundance gradient.

For $T_k$=100\,K, the [\OI]63\,$\mu$m line emission is subthermal ($T_{ex}$$<$T$_k$) which
means that the $^3P_1$ level ($n_1$) is much less populated than
in LTE (\textit{i.e.,} $\frac{n_1}{n_2}\ll\frac{3}{5}e^{-\frac{228}{100}}\simeq$$\frac{1}{16}$).
In fact we find that the  $^3P_0$--$^3P_1$ transition at 145\,$\mu$m is 
suprathermal (with $\frac{n_0}{n_1}\simeq\frac{1}{5}>\frac{1}{3}e^{-\frac{99}{100}}\simeq$$\frac{1}{8}$). 
This effect is produced by
the larger radiative life time of the upper $^3P_0$ level compared to the lower $^3P_1$
level (by a factor of $\sim$5). Even more, the lack of significant 
$^3P_0$-$^3P_1$ collisional de-excitation for \OI--H$_2$ collisions (neglecting those
with H) can invert the level populations 
($\frac{n_0}{n_1}>\frac{1}{3}$) leading to a   
[\OI]145\,$\mu$m maser line emission. High temperatures
($T_k$$\gtrsim$100\,K for a pure H$_2$ gas and $T_k$$\gtrsim$300\,K for a pure H gas) 
together with  relatively low $\tau_{63}$ opacities (or $n_\emr{H}$$<$10$^5$\,cm$^{-3}$)
are required
to obtain the level inversion, otherwise  line--trapping in the [\OI]63\,$\mu$m
line increases the upper $^3P_1$ level population again.
At such ``low'' gas densities ($n_\emr{H}< n_\emr{cr}$) the \textit{local} treatment of  \OI\, excitation 
overestimates (underestimates) the $n_1$ ($n_0$) level
populations, leading to stronger (weaker) [\OI]63\,$\mu$m (145\,$\mu$m)
emission as long as $\tau_{63}$$\gtrsim$1.  The error made
increases with $N$(\OI) 
(see  Elitzur \& Asensio--Ramos 2006).

The FIR continuum  toward the Horsehead is optically thin and
radiative pumping does not play a significant role in the 
\OI\,~excitation (it reduces $I_{63}$
by $\sim$1\,$\%$). 
However, in more extreme sources (with opaque continuum emission),
FIR pumping can reduce the $T_{ex}$
of the \OI\, $^3P_1$-$^3P_2$  transition ($T_{ex}$$\rightarrow T_d$) and
increase the $n_0/n_1$ level population ratio.
For the physical conditions treated here, the  
[\OI]63/145 ratio decreases by a factor of $\sim$2 if the dust emission becomes
optically thick at 63\,$\mu$m. Overall, neglecting  FIR pumping
and using \textit{local} approximations for the \OI\, excitation
overestimates the [\OI]63/145 ratio.

\subsection{Cooling rate and photoelectric heating efficiency}

Once the non-LTE  level populations have been accurately computed, 
one can determine the appropriate [\OI]63\,$\mu$m cooling rate per volume
unit,  $\Lambda_{63}$=$h\nu_{12}\,(n_2\,C_{21}$--$n_1\,C_{12})$
(where $C_{ij}$ are the collisional excitation and de-excitation rates in s$^{-1}$).
This relation takes into account that only line photons following 
collisional excitation and escaping the cloud are responsible of the gas cooling. 
We compute $\Lambda_{63}$$\simeq$5$\times$10$^{-21}$\,erg\,s$^{-1}$\,cm$^{-3}$ toward the 
PEAK position. 


Our observations and models are consistent
with the [\OI]63\,$\mu$m line arising from the outer layers ($A_V$$\simeq$0.1 and~0.5) 
of the Horsehead PDR ($G_0$$\simeq$100 and $n_\emr{H}$$\simeq$10$^4$\,cm$^{-3}$) where
the gas is warm ($\simeq$100\,K) and not all available hydrogen has converted 
into H$_2$ yet.
Together with
available [\CII]158\,$\mu$m observations 
($I_{158}\simeq(1-3)\times10^{-7}$~W\,m$^{-2}$\,sr$^{-1}$,
from \textit{KAO} at $\sim$55'' resolution; Zhou et al. 1993), we find that the luminosity of the 
[\CII]158\,$\mu$m~+~[\OI]63\,$\mu$m lines represents $\sim$1\%-2\% of the 
total FIR dust luminosity. This fraction corresponds to a high efficiency of the photoelectric
heating mechanism. 
In fact, due to the relatively low FUV  field but moderate
gas density in the PDR, the Horsehead is a good source to investigate the
grain populations that efficiently transform the incident FUV field into gas heating.

\textit{Herschel} 
will allow us to map almost all relevant FIR cooling lines at 
much higher spectral and angular resolutions.
Together with appropriate models of their excitation, 
it will then be possible to study the energy balance of 
interstellar clouds in greater detail. 

\clearpage

\acknowledgments
We thank the SPECPDR team for their contribution to the project,
A. Noriega-Crespo and R. Paladini for valuable discussions
on MIPS/SED data reduction and M. Gerin and the referee for helpful
comments on the manuscript.
JRG was supported by a \textit{Ram\'on y Cajal} research contract
from the Spanish MICINN and co-financed by the European Social Fund.

\clearpage



\end{document}